\documentclass[a4paper,fleqn,usenatbib]{mnras}

\usepackage[T1]{fontenc}
\usepackage{ae,aecompl}

\usepackage{graphicx}	
\usepackage{amsmath}	
\usepackage{amssymb}	

\title[Confirming the least massive Pleiades  members]{Confirming the least massive members of the Pleiades star cluster}

\author[M. R. Zapatero Osorio et al.]{
M. R. Zapatero Osorio,$^{1}$\thanks{E-mail: mosorio@cab.inta-csic.es}
V. J. S. B\'ejar,$^{2,3}$
N. Lodieu$^{2,3}$
and E. Manjavacas$^{4}$
\\
$^{1}$Centro de Astrobiolog\'ia (CSIC-INTA), Carretera Ajalvir km 4, E-28850 Torrej\'on de Ardoz, Madrid, Spain\\
$^{2}$Instituto de Astrof\'isica de Canarias, C/ V\'ia L\'actea s/n, E-38200 La Laguna, Tenerife, Spain\\
$^{3}$Universidad de La Laguna, Tenerife, Spain\\
$^{4}$Department of Astronomy/Steward Observatory, The University of Arizona, 933 N.~Cherry Avenue, Tucson, AZ, 85721, USA
}

\date{Accepted XXX. Received YYY; in original form ZZZ}

\pubyear{2017}

\begin{document}
\label{firstpage}
\pagerange{\pageref{firstpage}--\pageref{lastpage}}
\maketitle

\begin{abstract}
We present optical photometry ($i$- and $Z$-band) and low resolution spectroscopy (640--1015 nm) of very faint candidate members ($J = 20.2-21.2$ mag) of the Pleiades star cluster (120 Myr). The main goal is to address their cluster membership via photometric, astrometric, and spectroscopic studies, and to determine the properties of the least massive population of the cluster through the comparison of the data with younger and older spectral counterparts and state-of-the art model atmospheres. We confirm three bona-fide Pleiades members that have extremely red optical and infrared colors, effective temperatures of $\approx$1150 K and $\approx$1350 K, and masses in the interval 11--20 $M_{\rm Jup}$, and one additional likely member that shares the same motion as the cluster but does not appear to be as red as the other members with similar brightness. This latter object requires further near-infrared spectroscopy to fully address its membership in the Pleiades. The optical spectra of two bona-fide members were classified as L6--L7 and show features of K\,{\sc i}, a tentative detection of Cs\,{\sc i}, hydrides and water vapor with an intensity similar to high-gravity dwarfs of related classification despite their young age. The properties of the Pleiades L6--L7 members clearly indicate that very red colors of L dwarfs are not a direct evidence of ages younger than $\approx$100 Myr. We also report on the determination of the bolometric corrections for the coolest Pleiades members. These data can be used to interpret the observations of the atmospheres of exoplanets orbiting stars.
\end{abstract}




\section{Introduction}

Understanding the evolution and physical properties of brown dwarfs and isolated planetary-mass objects is key for the interpretation of atmospheric observations of unseen planets orbiting stars. To build the picture of the substellar evolutionary sequence, various groups are carrying out deep photometric surveys on different nearby star-forming regions and young star clusters with ages ranging from  $\approx$1 though 800 Myr \citep[e.g.,][]{dawson11,casewell12,casewell14,boudreault12,pena12,alves13,lodieu12,lodieu13,downes14,esplin17,luhman16,luhman17,schneider17,perez17}. With an age of 120\,$\pm$\,10 Myr \citep{basri96,martin98a,stauffer98}, a distance of 133.5\,$\pm$\,1.2 pc \citep{soderblom05,melis14,galli17}, and more than 1000 known members \citep{sarro14}, the Pleiades star cluster offers a unique opportunity to scrutinize brown dwarfs and planetary-mass objects of known (intermediate) age and solar metallicity. Previous attempts to identify Pleiades substellar objects with L and T types were reported by \citet{martin98b}, \citet{bihain06}, \citet{casewell07,casewell10}, and \citet{lodieu12}. Spectroscopic follow-up observations have confirmed L-type Pleiades brown dwarf members down to $\sim$L4 spectral type \citep{bihain10}, yet no T-type cluster object is unambiguously identified to date (see \citealt{casewell11,lucas13}), partly because these sources are intrinsically very dim and spectroscopic observations are challenging. 

Our group discovered a population of 19 faint, red Pleiades candidates in an area of 0.8 deg$^2$ whose proper motions are compatible with the distinctive one of the cluster \citep{osorio14a}. This survey covered about 3\%~of the total area of the cluster and extended the known Pleiades photometric sequence by about 2 mag fainter than any previous photometric and astrometric exploration carried out on the cluster. If corroborated as Pleiades members, the faintest candidates may have temperatures below the onset of methane absorption and masses below the deuterium burning mass limit of $\approx$13 M$_{\rm Jup}$. In a follow-up paper, \citet{osorio14b} reported on the near-infrared low resolution spectroscopy of six candidates in the interval $J = 17.5-20.3$ mag, the photometric $Z$-band detection of Calar Pleiades~21 ($Z = 22.91\pm0.15$ mag), and an upper limit on the $Z$-band magnitude of Calar Pleiades~25 ($Z \ge 23.35$ mag). These observations confirmed the cool nature of all sources with near-infrared spectral types determined in the range L2--L7. 

Here, we continue the follow-up optical observations of the six faintest Calar Pleiades\footnote{From now on, we will use the abridged name ``Calar''.} candidates from \citet{osorio14a}: Calar\,21--26. Their magnitudes are in the interval $J = 20.2-21.2$ mag, and their masses would range from 11 to 20 M$_{\rm Jup}$ if confirmed as cluster members. We report on new low resolution spectroscopy at visible wavelengths for Calar\,21 and~22, and the $Z$-band detections of Calar\,22, 23, 24 (also detected in the $i$-band), 25, and~26. Using the new optical data of this paper and the near-infrared spectra and near- and mid-infrared photometry published in \citet{osorio14a,osorio14b}, our objectives are twofold. We aim at robustly establishing their cluster membership via the confirmation of their proper motion measurements and their red spectrophotometric spectral energy distributions from optical to infrared wavelengths, and at discussing their photometric and spectroscopic properties in comparison with other young and old spectral counterparts. In Section~\ref{sec:observations}, we present the new imaging and spectroscopic observations. The analysis of the cluster membership based on astrometric, photometric and spectroscopic results is given in Section~\ref{sec:cmembership}. The properties of the confirmed least luminous Pleiades members are discussed in Section~\ref{sec:discussion}, where we also consider the spectral fitting of the data using state-of-the art model atmospheres and the calculation of bolometric luminosities and masses. Section~\ref{sec:conclusions} presents the conclusions of this paper. In an Appendix, we introduce new follow-up data of one high proper motion white dwarf discovered in \citet{osorio14a}.


\section{Observations}
\label{sec:observations}

All observations were acquired with the Optical System for Imaging and low-Intermediate-Resolution Integrated Spectroscopy (OSIRIS, \citealt{cepa00}), mounted on the Nasmyth focus of the 10.4-m Gran Telescopio de Canarias (GTC) on the Roque de los Muchachos Observatory. The detector of OSIRIS consists of two Marconi CCD42-82 (2048\,$\times$\,4096 pixels) with a small gap between them and with a pixel pitch of 15 $\mu$m. We only used the second detector since it is less affected by vignetting. The standard photometric and spectroscopic observing modes use binned pixels with a size of 0\farcs25822 on the sky \citep{sahlmann16}. Table~\ref{tab:obslog} provides the journal of the photometric and spectroscopic observations including the observing dates (Universal Time), exposure times, the airmass at the start of data acquisition, and the average seeing conditions. All nights were photometric; the typical seeing was of about 1\arcsec~or better.

In addition to the Calar candidate members, we also obtained OSIRIS data of the high proper motion source NPM\,03 \citep{osorio14a}; the acronym ``NPM'' stands for non-proper-motion candidate member of the Pleiades. This object lies close to Calar\,21 on the sky (both objects can be imaged simultaneously with OSIRIS) and has a proper motion twice as large as that of the cluster in the right ascension coordinate while its motion in the declination axis is similar to the Pleiades. We report on the observations of NPM\,03 in Tables~\ref{tab:obslog} and~\ref{tab:pm} and briefly discuss this particular object in the Appendix.

\begin{table}
\centering
\caption{Journal of photometric and spectroscopic observations with OSIRIS on the GTC. On-source exposure time is given as number of exposures multiplied by individual integration time.}
\label{tab:obslog}
\footnotesize
\begin{tabular}{p{1.08cm}p{1.56cm}p{0.99cm}p{0.8cm}p{0.7cm}p{0.7cm}}
\hline
Object & Obs$.$ date & Fil./Grat. & Exp$.$ & Airmass & Seeing \\
       &    (UT)     &            & (s)    &         &        \\
\hline
Calar 21 & 2014 Nov 24 & Sloan $z$  & 8$\times$90   & 1.08 & 0\farcs9 \\
Calar 21 & 2014 Nov 24 & R300R      & 4$\times$2700 & 1.04 & 0\farcs9 \\
Calar 22 & 2014 Nov 25 & Sloan $z$  & 36$\times$30  & 1.03 & 1\farcs0 \\
Calar 22 & 2015 Dec 11 & R300R      & 6$\times$2700 & 1.01 & 1\farcs0 \\
Calar 23 & 2015 Dec 12 & Sloan $z$  & 27$\times$60  & 1.39 & 0\farcs9 \\
Calar 24 & 2015 Dec 12 & Sloan $z$  & 27$\times$60  & 1.25 & 0\farcs7 \\
Calar 24 & 2015 Dec 12 & Sloan $i$  & 24$\times$90  & 1.19 & 0\farcs9 \\
Calar 25 & 2014 Nov 26 & Sloan $z$  & 36$\times$30  & 1.00 & 1\farcs0 \\
Calar 26 & 2015 Dec 12 & Sloan $z$  & 27$\times$60  & 1.46 & 0\farcs7 \\
\hline
NPM\,03  & 2014 Nov 24 & Sloan $z$  & 8$\times$90   & 1.08 & 0\farcs9 \\
NPM\,03  & 2016 Jan 11 & R300R      & 3$\times$1800 & 1.24 & 1\farcs2 \\
\hline
\end{tabular}
\end{table}

\begin{table*}
\centering
\caption{Photometry, proper motions, optical spectral types, and membership flag of Calar Pleiades candidates. The high proper motion source NPM03 is also listed. Calar 23 is not included because it is resolved using the OSIRIS data (see text). The $J$- and $K$-band magnitudes are taken from \citet{osorio14a}. $Z$, $J$, and $K$ are in the Vega system, $i$ magnitude is in the AB system.}
\label{tab:pm}
\begin{tabular}{lcccccccc} 
\hline
Object & $J$ & $K$ & $Z$  & $i$   & SpT & $\mu_\alpha\,{\rm cos}\,\delta$ & $\mu_\delta$     & Membership \\
       & (mag) & (mag)              & (mag) & (mag) &     & (mas\,yr$^{-1}$)                & (mas\,yr$^{-1}$) &            \\
\hline
Calar 21 & 20.23\,$\pm$\,0.09 & 17.73\,$\pm$\,0.03 & 22.98\,$\pm$\,0.09 &                    & L6--L7 & $+$22.0\,$\pm$\,6.0 & $-$37.7\,$\pm$\,4.5 & Y \\
Calar 22 & 20.29\,$\pm$\,0.07 & 17.83\,$\pm$\,0.13 & 22.98\,$\pm$\,0.10 &                    & L6--L7 & $+$11.9\,$\pm$\,9.0 & $-$56.6\,$\pm$\,7.7 & Y \\
Calar 24 & 20.65\,$\pm$\,0.13 & 18.83\,$\pm$\,0.33 & 23.09\,$\pm$\,0.08 & 25.80\,$\pm$\,0.19 &        & $+$12.8\,$\pm$\,7.0 & $-$44.5\,$\pm$\,5.7 & Y? \\
Calar 25 & 20.83\,$\pm$\,0.15 & 18.46\,$\pm$\,0.21 & 23.79\,$\pm$\,0.10 &                    &        & $+$23.4\,$\pm$\,8.2 & $-$47.9\,$\pm$\,6.8 & Y \\
Calar 26 & 21.15\,$\pm$\,0.14 & 19.40\,$\pm$\,0.50 & 23.71\,$\pm$\,0.10 &                    &        & $-$0.2\,$\pm$\,10.0 & $-$25.3\,$\pm$\,15.0 & N? \\
\hline                                        
NPM\,03  & 20.44\,$\pm$\,0.11 & 20.19\,$\pm$\,0.15 & 20.62\,$\pm$\,0.07 &                    & WD     & $+$44.8\,$\pm$\,6.0 & $-$34.9\,$\pm$\,4.5 & N \\
\hline
\end{tabular}
\end{table*}

\subsection{Photometry}
\label{sec:phot} 

We collected optical images of Calar\,21, 22, 23, 24, 25, and~26 using the Sloan $z$-band filter. Calar\,24 was also observed in the Sloan $i$-band in order to have an additional optical color that would help us understand whether it is an L- or a T-type source (Table~\ref{tab:obslog}). Total on-source exposure times ranged from 720 to 2160 s depending on the brightness of the target, observing filter, and seeing conditions. Individual images were acquired with short exposures (30--90 s) and with $\sim$10\arcsec~offsets, which allowed us to remove cosmic rays and the Earth's atmosphere contribution to the sky emission. Raw frames were bias subtracted and flat-field corrected. For debiasing, we used the overscan regions of the detector. For flat-fielding, we median combined all debiased images obtained during the night to produce a ``super-flat''. Individual frames were properly aligned and stacked to produce final deep $i$- and $z$-band images of the science targets and standard sources. 

All Calar objects were detected in the Sloan $z$-band filter with signal-to-noise (S/N) ratio of $\approx$\,7 and higher at the peak of the point-spread-functions (PSFs). Calar\,24 was detected with a weak S/N\,$\approx$\,4 in the very deep $i$-band exposure. Calar\,23, which was detected at the border of the $J$ and $H$ images in the proper motion study by \citet{osorio14a}, is resolved into an extense source in the OSIRIS data. Its full width at half maximum (FWHM) is about 1 pixel larger than the average seeing of point-like sources. We thus discarded Calar\,23 as a substellar candidate of the Pleiades cluster, and will not consider it further in this paper. All remaining Calar sources are unresolved in the $z$-band images and they remain as Pleiades candidate members. 

Aperture and PSF photometry was obtained using the ``phot'' package within the IRAF\footnote{IRAF is distributed by National Optical Astronomy Observatories, which is operated by the Association of Universities for Research in Astronomy, Inc., under contract with the National Science Foundation.} environment. We defined circular apertures of radii $\ge$4\,$\times$\,FWHM in size for deriving instrumental magnitudes. Photometric calibration of the $z$-band instrumental magnitudes of our targets was performed using the $Z$-band data of bright sources in the OSIRIS field of view provided by the UKIRT Infrared Deep Sky Survey (UKIDSS, \citealt{lawrence07}), which fully overlaps with the Pleiades. The Sloan magnitudes were converted into the UKIDSS $Z$-band system using the color terms appropriate for mid-to late-L types given by \citet{hewett06}. The $i$-band data of Calar\,24 was photometrically calibrated using OSIRIS observations of the field, young L5 dwarf 2MASS\,J03552337$+$1133437 \citep{reid06}, which was observed at similar air mass immediately after our target on the same night. We employed the Sloan magnitudes of the field L5 dwarf available from the Sloan Digital Sky Survey Data Release 11$^{\rm th}$ and 12$^{\rm th}$ \citep{alam15}. The final $i$- (AB system) and $Z$-band (Vega system) magnitudes are listed in Table~\ref{tab:pm}. 

The $Z$-band photometry of Calar\,21 reported here agrees with the measurement published by \citet{osorio14b} within less than 1 $\sigma$ the quoted error bars. This hints to long time scale photometric variability, if any, smaller than 0.15 mag in amplitude at these wavelengths. The actual $Z$ magnitude of Calar\,25 (Table~\ref{tab:pm}) also agrees with the upper limit reported in \citet{osorio14b}.

\subsection{Astrometry}
We used the OSIRIS images as the second epoch data to determine new proper motions for our targets following the methodology described in \citet{osorio14a}. This allowed us to confirm proper motions and identify sources with wrong astrometric measurements in the original paper. The time separation between the first epoch images ($J$-band, pixel size of 0\farcs3961, OMEGA' camera \citealt{osorio14a}) and the OSIRIS data ranges from 16.07 to 17.12 yr. This is 1.8--1.9 times longer than the time baseline of the first proper motion study, which would lead to more reliable astrometric determinations. In summary, the pixel coordinates of our targets and reference sources in the field were compared frame per frame and the coordinates transformations derived from the reference sources included third-order polynomials. Table~\ref{tab:pm} lists the proper motions of the Calar targets. Reported proper motion error bars correspond to the dispersion of the astrometric transformations, which included reference sources as bright and brighter than the targets. These errors are typically smaller by a factor of $\approx$\,1.3--2.0 than those reported in Table~1 by \citet{osorio14a}.

Calar\,26 is the faintest source in the $J$-band and is located at $\sim$\,2\arcsec~south from another source, which may add ``noise'' to its proper motion. To minimize this problem, we took advantage of the good seeing of the OSIRIS data (Table~\ref{tab:obslog}) and we derived the displacement of Calar\,26 using both $J$- and $H$-band images reported in \citet{osorio14a} that were acquired 17.09 and 9.09 yr before the OSIRIS image, respectively. The astrometric result given for Calar\,26 in Table~\ref{tab:pm} corresponds to the averaged value of the two determinations. The two measurements coincide to better than 1-$\sigma$ the quoted uncertainties.

\begin{figure}
	\includegraphics[width=\columnwidth]{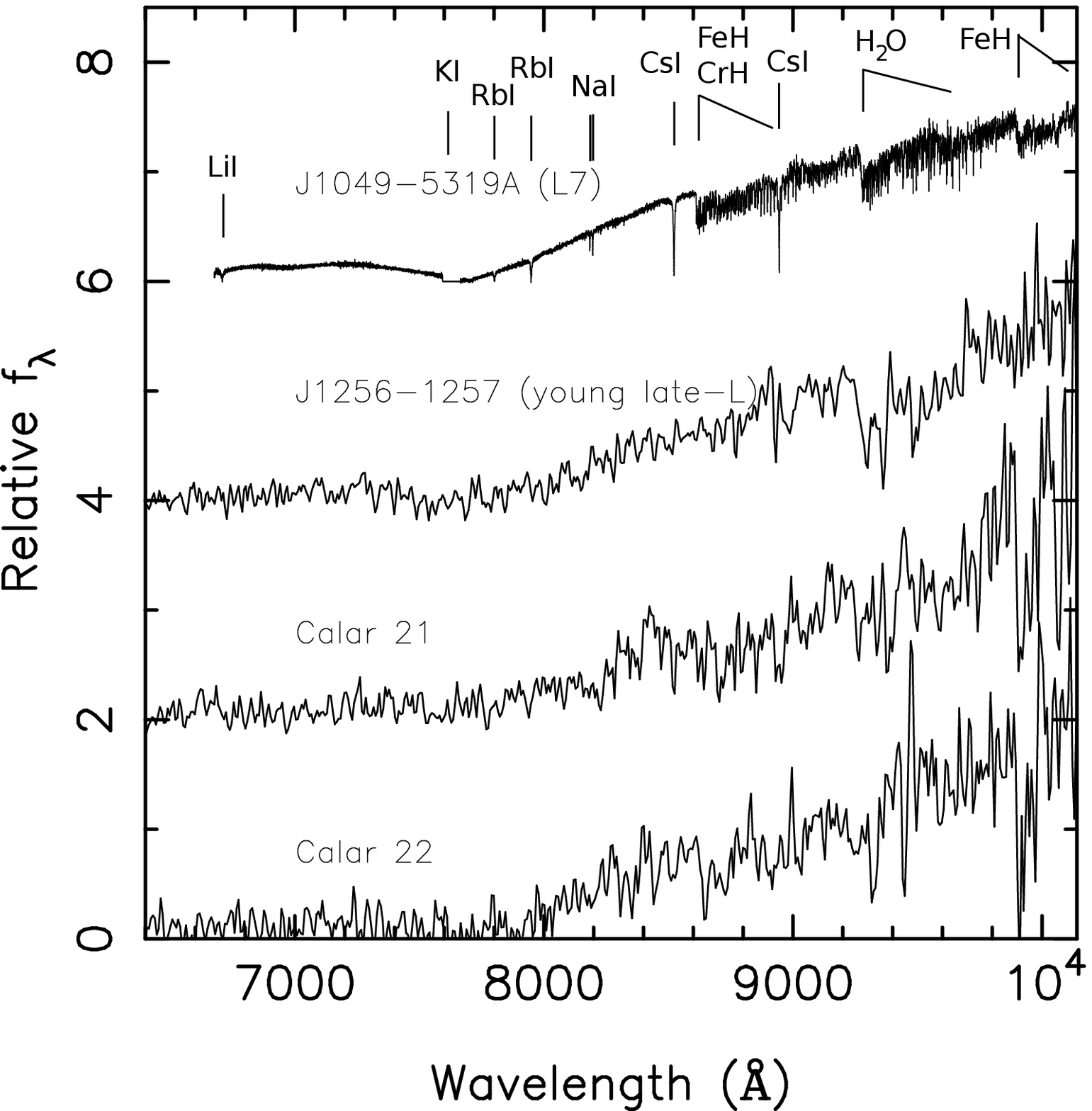}
    \caption{The OSIRIS optical spectra of Calar\,21 and~22 ($R \approx 270$) are shown together with the OSIRIS data of the young late-L dwarf J1256$-$1257b \citep[$R \approx 320$,][]{gauza15}. The top spectrum corresponds to the L8 high-gravity dwarf WISE\,J104915.57$-$531906.1A (Luhman~16A, \citealt{luhman13,kniazev13}) obtained with X-SHOOTER ($R \approx 11000$) and published in \citet{lodieu15}. It is depicted to illustrate the location of the most relevant features at these wavelengths. Spectra are normalized to unity at 9050--9150 \AA, and are shifted for clarity. Spectral features are identified after \citet{kirk99}.
}
\label{fig:spec}
\end{figure}

\subsection{Optical spectroscopy}
We obtained the optical spectra of Calar\,21 and~22, whose low-gravity nature was confirmed via near-infrared spectroscopy in \citet{osorio14b}, with two main objectives: the characterization of very red L dwarfs of well known age and metallicity at various wavelengths, and the determination of bolometric luminosities, which are fundamental for deriving masses. Optical spectra were acquired using OSIRIS, the R300R grism, and the long 1\arcsec~slit. Together with the 2$\times$ binned pixels of the OSIRIS detector, this instrumental configuration delivers low resolution spectra with a nominal dispersion of 7.8\,\AA\,pix$^{-1}$ or a resolving power of $R \sim 270$ at 8500\,\AA, and wavelength coverage between 5000 and 12000 \AA. The spectra were not acquired at parallactic angle since the targets and at least one bright reference star located at less than 1\arcmin~distance were simultaneously aligned on the slit. This observational strategy allowed us to register and combine individual short exposures (2700 s each) taken in AB cycles (nodding positions were typically separated by 8--10\arcsec~for proper subtraction of the Earth's sky emission contribution). Observations were executed at or near culmination (see Table~\ref{tab:obslog}) and targets were acquired on the slit using the $z$-band filter (passband of 8250--10000\,\AA). Because our Calar candidates have very little flux at blue wavelengths ($\le$6500\,\AA), we do not expect significant light losses through the narrow slit due to Earth's relative atmospheric refraction. The nodding observations were repeated to yield on-source integrations of 3 h (Calar\,21) and 4.5 h (Calar\,22).

Raw images were reduced with standard procedures including bias subtraction and flat-fielding within IRAF. The registered frames were stacked together before optimally extracting the spectra of the targets. A full wavelength solution from calibration lamps taken during the observing nights was applied to the spectra. The error associated with the fifth-order Legendre polynomial fit to the wavelength calibration is typically 10\%~of the nominal dispersion. We corrected the extracted spectra for instrumental response using data of the spectrophotometric standard white dwarf G191$-$B2B \citep{massey90}, which was observed on the same nights and with the same instrumental configuration as the targets. An order blocking filter blueward of 4500\,\AA~was used; however, there exists a second-order contribution redward of 9500\,\AA~(particularly important for blue objects), which was accounted for by observing the spectrophotometric standard star using the broad $z$-band filter and the same spectroscopic configuration as that of the science targets. To complete the data reduction, target spectra were divided by the standard star to remove the contribution of telluric absorption; the intrinsic features of the white dwarf were previously interpolated and its spectrum was normalized to the continuum before using it for division into the science data. The spectrophotometric standard star was observed a few hours before or after the targets; therefore, some telluric residuals may be present in the corrected spectra, particularly the strong O$_2$ band at 7605\,\AA.

The resulting reduced OSIRIS spectra of Pleiades candidates Calar\,21 and~22 are depicted in Figure~\ref{fig:spec}. The panel displays the wavelength range 6400--10150 \AA~because at shorter and longer wavelengths, the data have very poor S/N, and the instrumental response correction and wavelength calibration are not reliable at values beyond $\approx$\,1 $\mu$m. Together with the science targets, we also show the spectrum of the young, dusty L7 dwarf VHS\,J125601.92$-$125723.9b \citep{gauza15,stone16}, which was acquired with the same instrumentation and has comparable spectral resolution ($R \approx 320$) as our data (spectrum published in \citealt{gauza15}). The spectra of the Calar sources and VHS\,J125601.92$-$125723.9b (J1256$-$1257b) are alike with peaks of cross correlation functions of 0.96 (Calar~21 and J1256$-$1257b) and 0.92 (Calar~22 and J1256$-$1257b). The object J1256$-$1257b is believed to be 150--300 Myr old. The high-resolution spectrum ($R \approx 11000$) of the field late-L dwarf WISE\,J104915.57$-$531906.1A (known as Luhman~16A, L8 according to \citealt{luhman13} and \citealt{kniazev13}; L7.5 according to \citealt{faherty14}; and L6--L7.5 according to \citealt{lodieu15}) obtained with the X-SHOOTER instrument (see \citealt{lodieu15}) is also depicted in Figure~\ref{fig:spec} for illustrating the location of the most relevant atomic and molecular features at red-optical wavelengths. This field object likely has an age of $\approx$\,1 Gyr \citep{faherty14,lodieu15}.


\section{Cluster membership}
\label{sec:cmembership}

We revisited Calar\,21, 22, 24, 25, and~26 membership in the Pleiades cluster using the OSIRIS photometry, astrometry, and spectroscopy. The last column of Table~\ref{tab:pm} summarizes our conclusion on cluster membership for each candidate after this paper.

\subsection{Proper motions \label{sec:pm}}

\begin{figure}
	\includegraphics[width=\columnwidth]{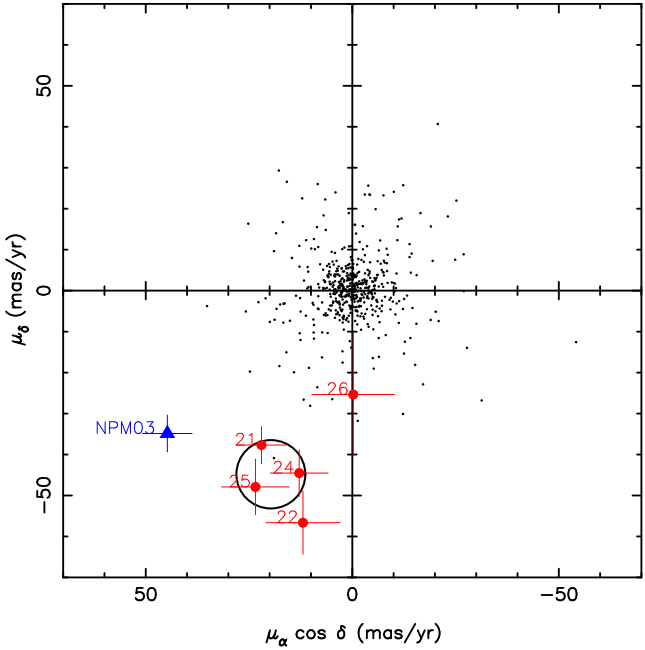}
    \caption{Proper motion diagram of the Pleiades candidates Calar\,21, 22, 24, 25, and~26 (red dots). The mean motion of the stellar cluster lies at the center [(19.71, $-$44.82) mas\,yr$^{-1}$, \citealt{loktin03}] of the black circle of radius of 8 mas\,yr$^{-1}$. The size of the circle corresponds to 1\,$\sigma$ astrometric error bar of sources with the same magnitudes as our targets. The tiny black dots around (0, 0) stand for the sample of objects contained within the field of view of the OSIRIS/GTC images and in common with the first epoch data. Our targets are labeled in red with their Calar identification number. The high proper motion source NPM\,03 (see the Appendix) is plotted as a blue triangle.
}
\label{fig:pm}
\end{figure}

Figure~\ref{fig:pm} shows the proper motion diagram resulting from the astrometric study presented here. The location of Pleiades members is given by the black circle, which is centered at the mean proper motion of the cluster \citep{loktin03} and has a radius equal to the average 1-$\sigma$ uncertainty of our study. As seen from Figure~\ref{fig:pm}, Calar\,21, 22, 24, and 25 lie close to the Pleiades motion, all except Calar\,26. The new proper motion of Calar\,26, although noisy, differs by more than 1-$\sigma$ the quoted uncertainties with respect to the astrometry of \citet{osorio14a}; however, both measurements are compatible at the 2.5 $\sigma$ level. Therefore, we consider that the membership of this object in the cluster has low probability. It is possible that the proper motion of Calar\,26 is affected by the proximity of another source unrelated to the cluster. We confirm the astrometry of the remaining Calar objects: it coincides within 1 $\sigma$ with the proper motions listed in \citet{osorio14a}. Calar\,21, 22, 24, and~25 are proper motion candidate members of the Pleiades.


\subsection{Spectral types  \label{sec:spt}}
\begin{figure}
	\includegraphics[width=\columnwidth]{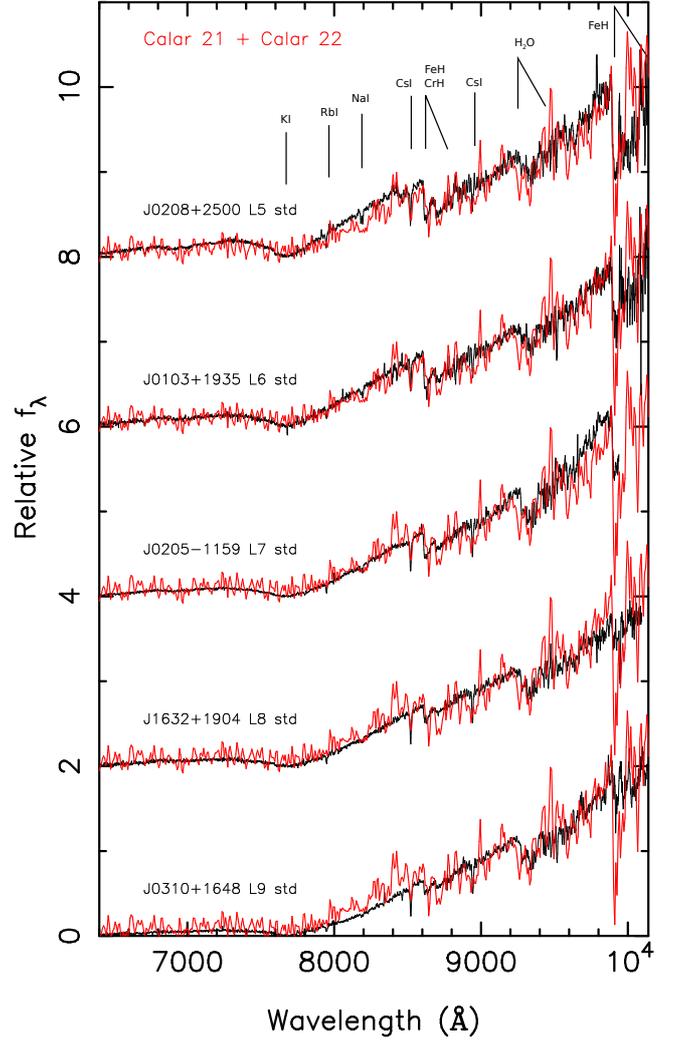}
    \caption{The combined OSIRIS spectrum of Calar\,21 and 22 (red) is compared to high-gravity L5--L9 primary spectral standards (black) for classification. All spectra are normalized to unity at 9050--9150 \AA. Data are shifted by a constant in the vertical axis for clarity. From top to bottom, dwarf standards are the following: 2MASS\,J02085499$+$2500488 (L5), 2MASS\,J01033203$+$1935361 (L6), DENIS-P\,J0205.4$-$1159 (L7), 2MASS\,J16322911$+$1904407 (L8), and 2MASS\,J03105986$+$1648155 (L9) from the catalogs of \citet{delfosse97}, and \citet{kirk99,kirk00}. The most relevant spectral features are indicated following \citet{kirk99}.
}
\label{fig:spt}
\end{figure}

We determined the spectral type of Calar\,21 and~22 at optical wavelengths by direct comparison with data from the literature. Both spectral standard dwarfs defined by \citet{kirk99,kirk00} and ``dusty'' dwarfs were employed for this task. To increase the S/N of the observed data, we combined the OSIRIS spectra of Calar\,21 and~22 into one single spectrum. This is justified by the similarity of Calar\,21 and~22 properties from visible to infrared wavelengths (see Figure~\ref{fig:spec}, Table~\ref{tab:pm}, and color-magnitude diagrams discussed in next Sections). The merged spectrum is shown against spectral standards ranging from L5 to L9 in Figure~\ref{fig:spt}. The optical spectra of the primary standard L dwarfs for spectral classification shown here were taken from \citet{delfosse97} and \citet{kirk99,kirk00}. These spectra\footnote{http://www.stsci.edu/\textasciitilde{}inr/ultracool.html} were collected with the Low Resolution Imaging Spectrograph (LRIS; \citealt{oke95}) at the 10 m W$.$ M$.$ Keck Observatory on Mauna Kea, Hawaii, and have a resolution of 9~\AA~over the wavelength range 6300--10100 \AA. This is a factor of $\approx$\,3 higher spectral resolution than our data; yet, it is acceptable for graphical purposes (Figure~\ref{fig:spt}). 

\begin{figure}
	\includegraphics[width=\columnwidth]{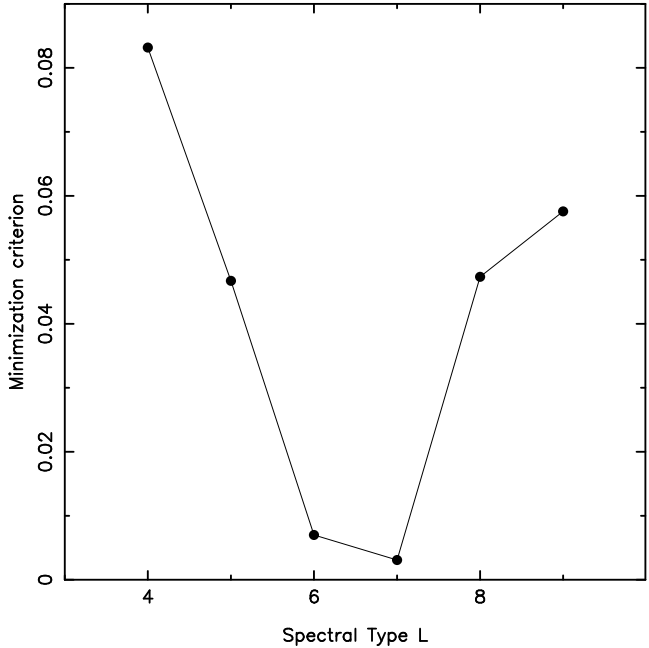}
    \caption{Figure of merit employed for the spectral classification of Calar\,21 and 22. The minimum is located between L6 and L7.}
\label{fig:minspt}
\end{figure}

We searched for the spectral primary standard that best resembles the Calar\,21$+$22 combined spectrum at ``all''  optical wavelengths. The following figure of merit (or minimization function) was defined:
\begin{equation}
{\rm merit} = \frac{1}{n} \sum_{i=1}^{n}|(f_{\rm obs}-f_{\rm std})_{\lambda_i}|
\end{equation} 
where $f_{\rm obs}$ and $f_{\rm std}$ correspond to the fluxes of the OSIRIS/GTC and LRIS/Keck spectra at wavelengths $\lambda_i$, and $n$ stands for the total number of wavelengths explored in the interval 6400--9880 \AA~at the matching wavelengths of our data. The spectra of the standard dwarfs were resampled to the resolving power of the OSIRIS data. The resulting figure of merit is illustrated in Figure~\ref{fig:minspt}. The minimum value of the figure corresponds to the best spectral resemblance. At optical wavelengths, Calar\,21 and~22 are classified as L6--L7 dwarfs. Note that the absolute minimum is located at L7 in Figure~\ref{fig:minspt}. However, the visual inspection of Figure~\ref{fig:spt} indicates that the L6 typing cannot be easily discarded. Other minimization functions, like the goodnes-of-fit statistic function described in \citet{cushing08}, show a broader minimum suggesting a spectral type in the interval L5--T0 (we used L and T standards and assigned equal weight to all wavelengths). 

This optical classification is consistent with the near-infrared spectral typing of \citet{osorio14b}: L7\,$\pm$\,1 (Calar\,21) and L/T (Calar\,22), which was assigned based on the comparison of the infrared $H$ and $K$-band data to spectra of young, dusty L dwarfs. However, the optical types appear to be slightly earlier than the near-infrared ones; this signature is found among young ultracool dwarfs and has been mentioned by, e.g., \citet{allers13}, \citet{osorio17}, and \citet{lodieu17}. Particularly, \citet{allers13} discussed that their classification of young L dwarfs of the field and star moving groups shows an offset of $\approx$1 subtype between infrared and optical spectral types. This phenomenon might be related to the low gravity atmospheres of our targets and the impact of dust absorption at short wavelengths. Calar\,21 and~22 have optical and near-infrared spectral types expected for Pleiades members with magnitudes around $K = 18$ mag; they follow the spectral type---magnitude relationships of \citet{dupuy12}, \citet{faherty16}, and \citet{liu16}. In addition, \citet{osorio14b} discussed the low gravity features observed in the near-infrared spectra of Calar\,21 and~22, and concluded that these two objects are spectroscopically confirmed members of the cluster.

\subsection{Color-magnitude diagrams \label{sec:cmd}}

\begin{figure}
	\includegraphics[width=\columnwidth]{zk.ps}
	\includegraphics[width=\columnwidth]{zj.ps}
    \caption{Color-magnitude diagrams of Pleiades Calar candidates (red dots) using observed MKO photometry. The red asterisk corresponds to the likely non-member of the cluster. The Calar sources are labeled in red with their identification number. Previously known cluster members \citep{bihain10,osorio14a} are depicted with black dots; the arrows indicate $Z$-band upper limit magnitudes. The BT-Settl 120-Myr, solar metallicity isochrone \citep{allard14,baraffe15} is plotted as a solid line. Predicted masses and effective temperatures are labeled on the right and left sides of the diagrams. The companion J1256$-$1257b, which has a likely age in the interval 150--300 Myr \citep{gauza15}, is shown in blue color; it was taken to the distance of the Pleiades cluster. 
}
\label{fig:zk}
\end{figure}

\begin{figure}
	\includegraphics[width=\columnwidth]{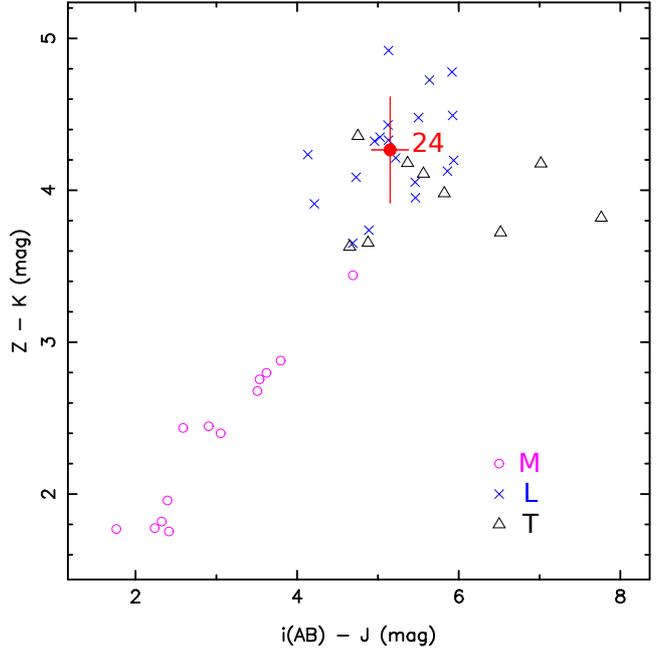}
    \caption{Color-color diagram to illustrate the ($i-J, Z-K$) indices of Calar\,24 (red dot) with respect to field M (open circles), L (crosses), and T (open triangles) dwarfs listed in \citet{hewett06}. L and early-T sources show significant overlap, thus impeding the unambiguous spectrophotometric classification of Calar\,24 with current photometric data. 
}
\label{fig:colorcolor}
\end{figure}

We used the $i$- and $Z$-band photometry to assess the cluster membership of the Calar sources. In Figure~\ref{fig:zk}, we show the $K$ versus $Z-K$ (top panel) and $J$ versus $Z-J$ (bottom panel) color-magnitude diagrams including previously known, confirmed Pleiades members with spectral types mid-M through mid-L, and our candidates. All magnitudes are given in the same photometric system allowing a direct comparison. Calar\,21, 22, and~25 nicely extrapolates the cluster sequence towards fainter $J$- and $K$-band magnitudes and redder $Z-K$ colors. As illustrated in the bottom panel, the $Z-J$ colors of the Calar candidates do not appear redder than those of brighter cluster members. Yet, Calar\,25 is the reddest object in our sample; it has $Z-J = 3.19 \pm 0.20$ and  $Z-K = 5.33 \pm 0.23$ mag. These indices are comparable to other very red, dusty late-L dwarfs like J1256$-$1257b, which is also plotted in Figure~\ref{fig:zk} (J1256$-$1257b was taken to the distance of the Pleiades by using the new trigonometric parallax determination of 15.8$^{+1.0}_{-0.8}$ pc determined by Zapatero Osorio et al$.$ 2017, in preparation). 

Calar\,24, although dimmer in $K$-band than Calar\,25, has bluer $Z-K$ and $J-K$ colors and deviates by about 1 and 0.5 mag, respectively, from the increasingly red sequence of the cluster (see Figures~\ref{fig:zk} and~\ref{fig:hr}). Whether this is evidence of the photometric non-membership of Calar\,24 in the Pleiades remains to be proved. On the one hand, Calar\,26, whose proper motion indicates it is not a likely cluster member (Section~\ref{sec:pm}), has $Z-K$ and $J-K$ colors similar to Calar\,24. This may suggest that Calar\,24 is a contaminant in the proper motion survey. On the other hand, in a recent work by \citet{vos17}, it is discussed that the viewing angle influences the spectral and photometric appearance of ultracool, dusty dwarfs: objects viewed equator-on appear redder than objects viewed at lower inclinations. This would imply a large dispersion in the colors and magnitudes of coeval, equal mass objects. In addition, the turn-over sequence that is expected at optical and near-infrared color-magnitude diagrams and is caused by the onset of methane absorption and/or dust settlement below the photosphere in ultracool atmospheres, should occur at magnitudes near the position of Calar\,25 in the color-magnitude diagrams. Calar\,24 occupies a position in Figures~\ref{fig:zk} and~\ref{fig:hr} well matched by the expected location of the ``methane sequence'' in the Pleiades. Furthermore, Calar\,24 shows $i-J$ and $i-Z$ colors similar to SDSS J085834.42$+$325627.6, which is a T0--T1 dwarf that displays redder-than-normal colors for its given spectral type \citep{faherty09,gagne14}. However, the kinematic study of \citet{gagne14} indicates that this T dwarf does not belong to any well-defined young star moving group (TW Hydrae, $\beta$ Pictoris, Tucana-Horologium, Carina, Columba, Argus, and AB Doradus) and has a low probability of being a member of the young field. In color-color diagrams, like the one illustrated in Figure~\ref{fig:colorcolor}, Calar\,24 does not stand out as a T dwarf; early-Ts and Ls have some overlapping properties in various colors. At present, and with current astrometric and photometric data, Calar\,24 cannot be rejected as a Pleiades candidate member, but it cannot be confirmed either. The cluster membership flag of Calar\,24 in the last column of Table~\ref{tab:pm} includes a ``?'' to remind that this object's membership status is still open. Spectroscopic observations at near-infrared wavelengths are required to explore the presence of methane absorption and other features typical of cool, low gravity atmospheres that would unambiguously confirm/reject Calar\,24 as a Pleiades member and would contribute to characterize the ``methane sequence'' of the cluster.

\section{Discussion}
\label{sec:discussion}

\subsection{Spectral properties   \label{sec:specprop}}
At the spectral resolution and quality of our data, the following spectroscopic features are seen in the optical spectrum of Calar\,21$+$22 (Figure~\ref{fig:spt}): the broad line of the resonance doublet of neutral potassium, K\,{\sc i}, centered at 7665 and 7699 \AA, molecular absorptions due to metallic hydrides FeH ($\approx$\,8692, 9896 \AA) and CrH ($\approx$\,8611 \AA), and absorption due to water vapor ($\approx$\,9300 \AA). All of them show intensities comparable to those of the late-L objects in the field. The K\,{\sc i} absorption is strong and defines the pseudocontinum of the optical energy distribution \citep{pavlenko00,allard01}. Although metallic hydrides are clearly present, molecular absorption due to oxides (TiO and VO) is not obvious, except for a weak feature at 8400--8600 (flat flux) that might be caused by TiO. This adds support to our spectral classification (VO and TiO have vanished by mid-L types at the typical atmospheric gravities of the field, \citealt{martin99,kirk99}). 

\begin{figure}
	\includegraphics[width=\columnwidth]{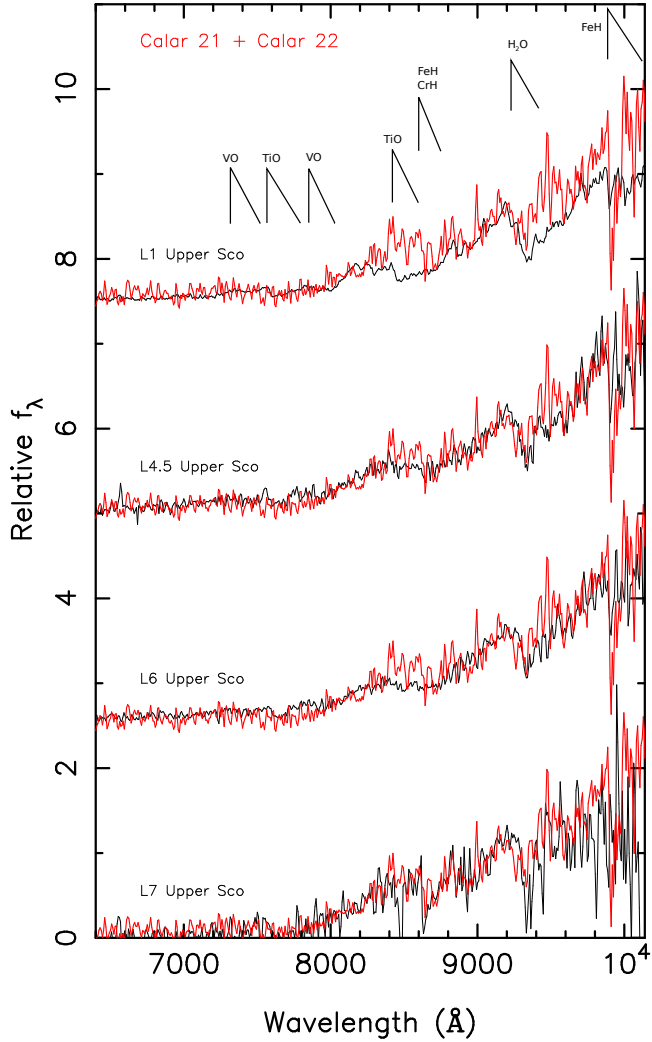}
    \caption{The OSIRIS merged spectrum of Calar\,21 and~22 (red) is compared to L-type spectra of Upper Scorpius members (black). All data were obtained with the same instrument, are normalized to unity at $\approx$9100--9150 \AA, and are vertically shifted by a constant for clarity. Some molecular features are indicated after \citet{kirk99}. }
\label{fig:upsco}
\end{figure}

A closer inspection of FeH features in Figure~\ref{fig:spt} reveals that they are slightly stronger in the spectra of the Pleiades objects than in the older, field L6--L7 standards. Furthermore, Calar\,21 and~22 present stronger FeH absorption at 9896 \AA~than the young J1256$-$1257b (see Figure~\ref{fig:spec}). This object's age is kinematically estimated at about 300 Myr (Zapatero Osorio et al$.$ 2017, in preparation), i.e., older than the Pleiades, which is consistent with the lithium depletion observed in the primary binary brown dwarf of the system \citep{gauza15} and with the age estimate of \citet{rich16}. Under the assumption that all sources shown in Figures~\ref{fig:spec} and~\ref{fig:spt} have identical metallicity, FeH absorption at optical wavelengths appears to be intense in late-L Pleiades objects. This behavior is opposite to the one described by \citet{mcgovern04} and \citet{martin17}: low gravity dwarfs show weaker hydride absorption. However, these authors focused on the $J$-band spectra of late-M and early-L dwarfs. \citet{lodieu17} show spectra of Upper Scorpius candidate members spanning the spectral range L0--L7. FeH is also seen in weak absorption with respect to the high gravity dwarfs of similar classification. If not due to gravity (atmospheric pressure), other explanations to account for the distinct FeH intensities are that the Pleiades cluster has different metallicity from the comparison objects, poor S/N of our data and bad telluric correction, and/or J1256$-$1257b and Calar~21 and~22 have distinct spectral types (FeH becomes weaker with decreasing temperature). The former scenario is less likely, since Pleiades F- and G-type stars have nearly solar Fe abundance \citep{soderblom09,takeda17}. The low S/N and unequal classifications are more plausible: J1215$-$1257b has a $K$-band absolute magnitude $\approx$\,1.5 mag fainter than those of Calar\,21 and~22 (see Section~\ref{sec:cmd}). The spectral classification of very faint (low luminosity), dusty dwarfs is not solidly defined yet \citep{allers13,faherty16,liu16}. By studying a sequence of faint and red sources that have the same age, metallicity and distance but different magnitudes, i.e., members of a star cluster, one can establish the sequence of spectral types in a consistent manner.

There is a tentative detection of neutral cesium at 8521 \AA~in the merged spectrum of Figure~\ref{fig:spt}. It is the strongest Cs\,{\sc i} atomic line at visible wavelengths. We measured a pseudo-equivalent width (pEW) of 8.6\,$\pm$\,0.5 \AA~by adopting two pseudo-continuum points at 8500 and 8545 \AA~around the line center. This value is compatible with the high measurements of field late-L dwarfs reported by \citet{kirk99,kirk00}, \citet{burgasser03}, \citet{lodieu15}, \citet{manjavacas16}, and references therein. Alkali lines are very sensitive to both temperature and gravity. At low gravity atmospheres and cool temperatures down to $\approx$1000 K, we would expect weaker atomic lines. Our Cs\,{\sc i} $\lambda$8521 \AA~pEW may be affected by the very low spectral resolution of our data and the poor S/N. We imposed an upper limit of 3.5 \AA~on Cs\,{\sc i} $\lambda$8943 \AA, and 6 \AA~on the pEW of Rb\,{\sc i} $\lambda$7948 \AA. This is slightly below the actual pEWs corresponding to field L7 dwarfs, which is consistent with low pressure atmospheres.

Figure~\ref{fig:upsco} displays various L-type optical spectra of the very young Upper Scorpius association (UpSco, $\sim$5--10 Myr, \citealt{preibisch02,song12,pecaut12,pecaut16}) obtained with the same instrumental setup as our data. Each spectrum was fabricated by averaging data of two different UpSco members with the same spectral classification that were recently published by \citet{lodieu17}. We adopted the ``near-infrared'' spectral classification scheme proposed by these authors. The global morphology of the Pleiades spectrum appears very similar to UpSco L7 dwarfs in Figure~\ref{fig:upsco}. FeH $\lambda$8692 \AA~absorption intensity is slightly stronger in the Pleiades. At earlier spectral types, and as shown in Figure~\ref{fig:upsco}, oxides (TiO and VO) are still present in the optical spectra of very young dwarfs. To study the dependence of spectroscopic features on gravity at late-L types, data obtained to a higher S/N are required.



\subsection{The Pleiades sequence \label{sec:sequence}}
\begin{figure}
	\includegraphics[width=\columnwidth]{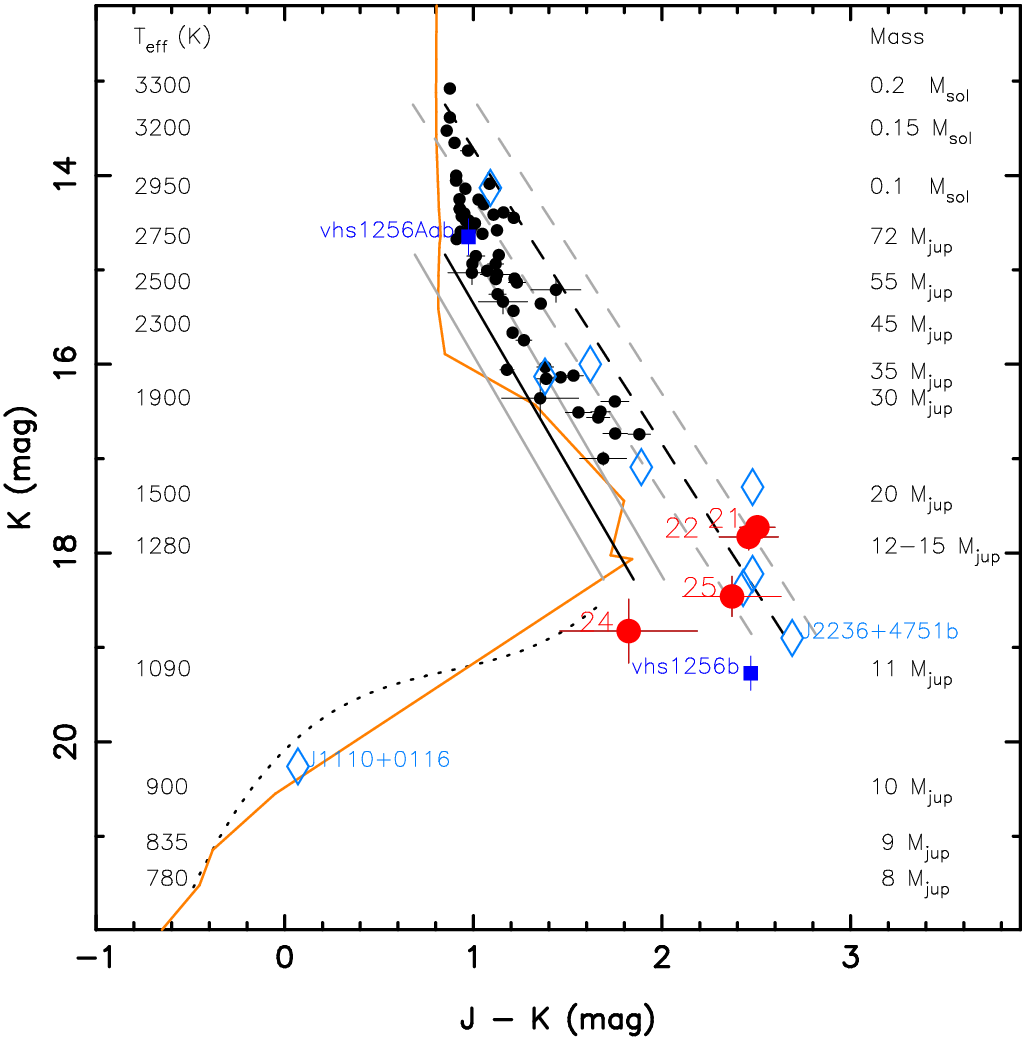}
\caption{Color-magnitude diagram of Pleiades members (dots). Red dots stand for the Calar sources (labeled with their identification number) with high cluster membership probability (this excludes Calar 26). Black dots stand for previously known mid-M through mid-L Pleiades members \citep{bihain10,osorio14a}. L- and T-type members of the 120-Myr AB Doradus moving group that have trigonometric distances (see \citealt{liu16,faherty16}) are plotted as open blue symbols: the L7 dwarf J2236$+$4751b \citep{bowler17} in the only AB\,Dor candidate member that has no parallax to date. The system J1256$-$1257 \citep{gauza15}, which has a likely age in the interval 150--300 Myr, is shown with solid blue squares. The black solid and dotted lines show the fits for field, high-gravity L- and T-type dwarfs obtained by \citet{liu16} and \citet{filippazzo15}, respectively. The sequence of low-gravity L dwarfs defined by \citet{liu16} is illustrated with a black dashed line. The gray solid and dashed lines show the 1\,$\sigma$ scatter about their corresponding linear fits. The BT-Settl solar metallicity, 120-Myr isochrone is shown with a brown solid line. Masses and effective temperatures predicted by this evolutionary model are given on both sides of the diagram. All AB\,Dor members, J1256$-$1257, the field low- and high-gravity sequences, and the isochrone were taken to the distance of the Pleiades cluster.
}
\label{fig:hr}
\end{figure}

Figure~\ref{fig:hr} depicts the sequence of Pleiades members contrasted with that of the AB Doradus moving group, which has an age (150 Myr, \citealt{bell15}) very close to the Pleiades. Actually, some authors argue that the Pleiades star cluster and AB Doradus moving group are likely coeval and share a common origin (e.g., \citealt[and references therein]{ortega07}). Only confirmed AB Doradus members with trigonometric distances are shown (see \citealt{faherty16} and \citealt{liu16} for details). The companion 2MASS J22362452$+$4751425b (J2236$+$4751b) to a late-K dwarf likely belonging to the AB Doradus moving group \citep{bowler17} is also included in the diagram, although it has no parallax available to date. \citet{bowler17} discuss that J2236$+$4751b may define the ``elbow'' of the AB Doradus photometric sequence separating L dwarfs from the cooler T dwarfs. All AB Doradus candidate members have been taken to the Pleiades distance for a proper comparison. As illustrated in Figure~\ref{fig:hr}, both Pleiades and AB Doradus sequences of M and L dwarfs nicely overlap, including the reddest $J-K$ colors. The exact location of the L/T transition (``elbow'') in the Pleiades is still pending the spectroscopic confirmation of the presence of methane in Calar\,24 (see above). Furthermore, these observations would be helpful to constrain the theoretical predictions of \citet{marley12}. These authors provided a toy model for the definition of the L/T transition where the effective temperature ($T_{\rm eff}$) range of the transition is made gravity dependent. The toy model was ``calibrated'' against the observations of high gravity dwarfs and the planets around HR\,8799 \citep{marois08,marois10}. According to \citet{marley12}, the ``elbow'' of the Pleiades sequence is expected at 900--1000 K or 10 M$_{\rm Jup}$. This is $\approx$1.5 mag fainter in $K$-band than the position of Calar\,24 or J2236$+$4751b in Figure~\ref{fig:hr}. \citet{rajan17} have adopted these models as one possibility to explain the red spectral energy distribution of the young T dwarf 51\,Eridani~b \citep{macintosh15}, which orbits 51\,Eri A, an F0IV star that is part of the $\beta$~Pictoris moving group. 

The Pleiades sequence is also compared to the tracks delineated by high gravity M, L, and T dwarfs in Figure~\ref{fig:hr}. The mean location and dispersion of field M6--L8 sources is adopted from \citet{liu16}; as for the field T dwarf sequence, we adopted \citet{filippazzo15}. All photometry is taken to the distance of the Pleiades cluster. It becomes apparent from the Figure that the Pleiades sequence of M, early-L, and mid-L dwarfs is overluminous in the $K$ versus $J-K$ diagram with respect to the field by about 0.5--1 mag ($K$), as expected for its youth: although at a slower pace than at much younger ages, Pleiades brown dwarfs and planetary-mass objects are still collapsing under their own gravity \citep{burrows93,baraffe02,saumon08}. As an example, dwarfs with masses of 13 and 50 M$_{\rm Jup}$ have sizes larger than Jupiter at 120 Myr and will have their radii reduced by $\approx$27\,\%~and $\approx$65\,\%, respectively, between the age of the Pleiades and 5 Gyr. 

The faintest and reddest Pleiades members (e.g., Calar\,21, 22, and~25) are significantly redder and slightly brighter ($\sim$0--0.5 mag) than their field, high gravity late-L counterparts in the $K$-band (Figure~\ref{fig:hr}, Table~\ref{tab:absmag}). Since one possible explanation for their reddish appearance lies on the presence of very thick clouds of dust, non-equilibrium chemistry, and rapid vertical mixing in the atmospheres \citep[and references therein]{madhusudhan11,barman11,skemer14}, it may be feasible that absorption/scattering of condensates is ocurring even at long wavelengths, thus yielding ``obscured'' $Z$, $J$, $H$, and $K$ magnitudes: the shorter the wavelength, the more affected by dust absorption. This would contribute to explain the apparent lack of brightness at these wavelengths with respect to the field. This fact was already pointed out by \citet{marocco14}, who stated that the diversity in near-infrared colors and spectra seen in late L dwarfs could be due to differences in the optical thickness of the dust cloud deck. This same argument is also employed to account for the observed photometric modulations in ultracool dwarfs \citep[e.g.,][and references therein]{yang15,lew16}. For other theoretical interpretations of the very red colors of L dwarfs, see \citet{tremblin16}.

In Figure~\ref{fig:hr}, we also included the ``very low gravity'' sequence of M6--L7 dwarfs defined by \citet[their Figure 14]{liu16}. These authors used the criteria of \citet{allers13} for classifying the gravity of the near-infrared spectra of young sources in the field and in stellar moving groups. The ``very low gravity'' flag roughly corresponds to the optical classification of $\gamma$ \citep{cruz09} and to ages of $\le$30 Myr (i.e., $\ge$4 times younger than the Pleiades). Whereas the very low gravity sequence runs above the Pleiades between mid-M and mid-L dwarfs in Figure~\ref{fig:hr}, which is consistent with two distinct ages, we caution that both sequences steadily converge at the late-L types (or faintest magnitudes). As discussed by \citet{bihain06} and \citet{luhman12}, this feature might be a natural consequence of the early evolution of brown dwarfs and planetary-mass objects: below $\sim$1500 K all low mass objects are predicted to have similar radii, even the younger ones \citep{burrows97,chabrier00a}. However, we remark that atmospheric gravity is not as similar. For instance, a 1300 K dwarf has a surface gravity of log\,$g$\,=4.45 [cm\,s$^{-2}$] at the age of 120 Myr, 4.06 dex at 20 Myr, and 5.10 dex at 1 Gyr \citep{chabrier00a}. Between the youngest and oldest ages, there is a difference of one order of magnitude in the surface gravity for a given temperature. Between 20 and 120 Myr, the gravity parameter differs by a factor of 2.5. Yet, this induces noticeable changes in ultracool spectra according to theoretical model atmospheres (e.g., \citealt{allard01,marley12}; see also Section~\ref{sec:models}). Calar\,21, 22, and~25 (120 Mr), together with the AB Doradus members (150 Myr) with similar absolute magnitude and colors, and J1256$-$1257b (150--300 Myr), strongly demonstrate that intermediate-age late-L dwarfs can posses extreme atmospheric properties typically attributed to much younger objects (e.g., 2MASS J1207334$-$393254b, \citealt{chauvin04}; PSO J318.5338$-$22.8603, \citealt{liu13}). See also the discussion by \citet{bowler17}. We caution that extremely red colors of L dwarfs are not a direct evidence of ages younger than $\approx$100 Myr.



\begin{table}
\centering
\caption{Absolute magnitudes for L7 dwarfs of different gravity.}
\label{tab:absmag}
\resizebox{\columnwidth}{!}{
\begin{tabular}{cccccl}
\hline
SpT & M($Z$) & M($J$) & M($H$) & M($K$) & Reference\\
    & (mag)  & (mag)  & (mag)  & (mag)  & \\
\hline
L7 Pleiades & $\approx$17.32 & $\approx$14.62 & $\approx$13.20 & $\approx$12.12 & This paper\\
\hline
L7 VL-G     &                & 15.90$\pm$0.72 & 14.54$\pm$0.64 & 13.30$\pm$0.56 & \citet{liu16}\\
L7 INT-G    &                & 15.07$\pm$0.50 & 13.61$\pm$0.43 & 12.59$\pm$0.39 & \citet{liu16}\\
\hline
L7 field    &                & 14.32$\pm$0.37 & 13.43$\pm$0.33 & 12.68$\pm$0.31 & \citet{liu16}\\
L7 field    &                & 14.46$\pm$0.40 & 13.35$\pm$0.39 & 12.62$\pm$0.54 & \citet{faherty16}\\ 
L7 field    & 16.72$\pm$0.52 &                &                &                & \citet{hewett06}\\
\hline
\end{tabular}
}
\end{table}

\begin{figure}
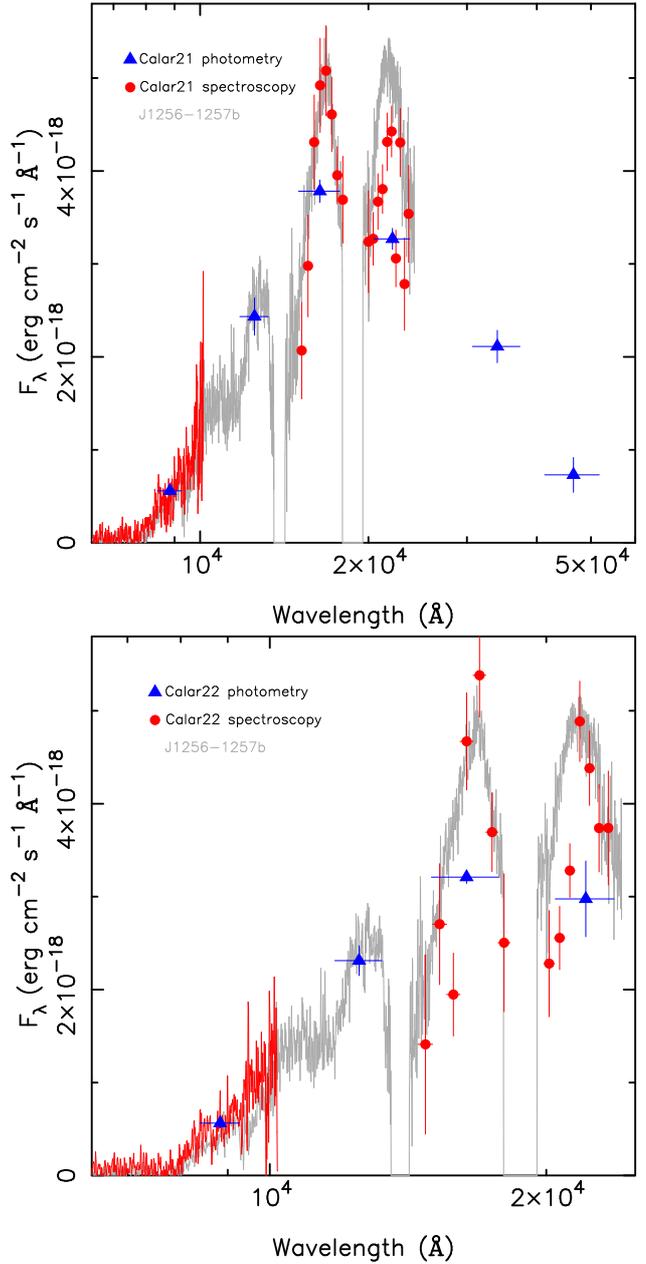

	\includegraphics[width=0.98\columnwidth]{calar21_sed.ps}
	\includegraphics[width=0.98\columnwidth]{calar22_sed.ps}
    \caption{Observed spectrophotometric energy distribution of Calar\,21 (top) and~22 (bottom) using optical, near- and mid-infrared data. The optical (this paper) and near-infrared spectra \citep{osorio14b} are shown in red color, while the photometry is plotted as blue solid triangles. For comparison purposes, the spectrum of J1256$-$1257b \citep{gauza15} is shown with a gray line; it is normalized to the total $J$-band flux of Calar\,21 and~22. The horizontal error bars associated with the photometry indicate the passband of the filters. The $x$-axis (wavelength) is shown in logarithmic scale.
}
\label{fig:seds}
\end{figure}

\begin{figure*}
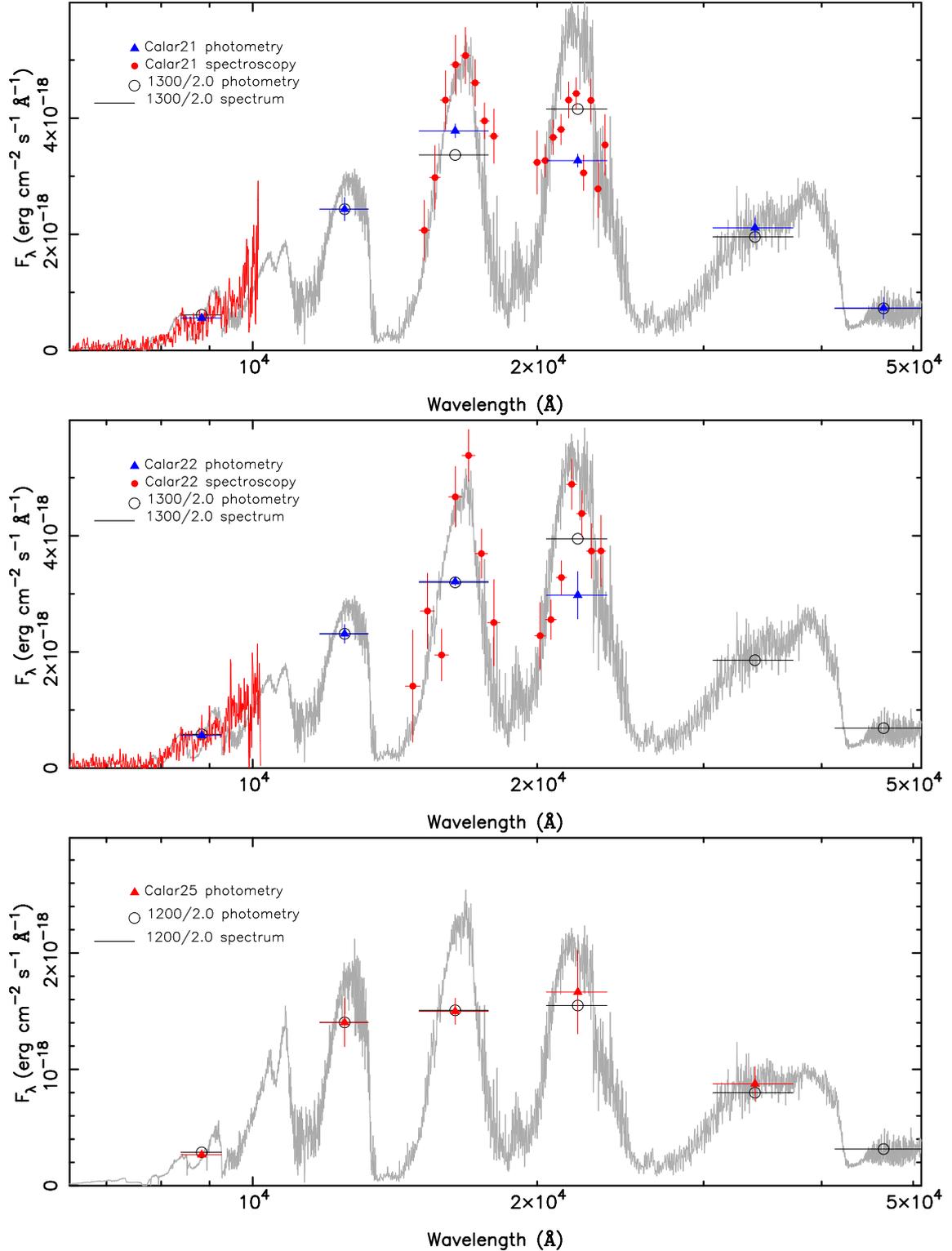

	\includegraphics[width=0.8\columnwidth, angle=-90]{calar21_model.ps}
	\includegraphics[width=0.8\columnwidth, angle=-90]{calar22_model.ps}
	\includegraphics[width=0.8\columnwidth, angle=-90]{calar25_model.ps}
    \caption{Observed spectrophotometric energy distribution of Calar\,21 (top), Calar\,22 (middle), and Calar\,25 (bottom). Synthetic BT-Settl spectra are depicted with gray solid lines. Theoretical photometry as integrated from the displayed models are plotted as open black circles. The models are normalized to the $J$-band fluxes of each Calar source. The horizontal error bars associated with the photometry indicate the passband of the filters (from blue to red wavelengths: $Z$, $J$, $H$, $K$, $W1$, and $W2$). The $x$-axis (wavelength) is shown in logarithmic scale. The models shown correspond to solar metallicity, $T_{\rm eff}$\,=\,1300 and 1200 K, and log\,$g$\,=\,2.0 dex.
}
\label{fig:models}
\end{figure*}

In the Pleiades, L7 type dwarfs have $Z \approx 22.95$, $J \approx 20.25$, $H \approx 18.83$, and $K \approx 17.75$ mag, corresponding to the absolute magnitudes at the age of the cluster (120 Myr) listed in Table~\ref{tab:absmag}. In the Table, and for comparison purposes, we also included the $ZJHK$ absolute magnitudes representing the high gravity dwarfs of related classification provided by \citet[Table~10]{hewett06}, \citet[Table~10]{liu16}, and \citet[Table~19]{faherty16}. These values stand for L7 dwarfs that typically have an older age than the Pleiades, and therefore, higher gravity atmospheres. All $ZJHK$ data are in the Mauna Kea Observatories (MKO) photometric system for a direct comparison \citep{simons02,tokunaga02}. The 2MASS $JHK$ magnitudes of \citet{faherty16} were transformed into the MKO system using the equations of \citet{stephens04}. We used the mean $Y$ magnitudes of \citet{liu16} and the $Z-Y$ colors of \citet{hewett06} to fill the $Z$ datum of field L7 dwarfs in Table~\ref{tab:absmag}. Pleiades late-L dwarfs are dimmer by $\approx$0.6 ($Z$) and $\approx$0.25 ($J$) mag than their field spectral counterparts, but are brighter in $H$ and $K$ by similar amounts, a signpost of their extremely red colors. This confirms the ``pivotal'' nature of the SEDs of young red L-type objects mentioned by \citet{filippazzo15}. Since one expects that young, inflated dwarfs are overluminous at all wavelengths with respect to the field, this supports evidence of significant scattering of light in dusty extended ultracool atmospheres.


\subsection{Spectral energy distribution \label{sec:models}}
We built the spectral energy distribution (SED) of Calar\,21, 22, and~25 from optical to infrared wavelengths by combining data presented here with data from the discovery paper and \citet{osorio14b}. The goal is a better characterization of the least massive population of the Pleiades. The $Z$ and $HK$ photometry was used to flux calibrate the optical and near-infrared spectra of Calar\,21 and~22 by integrating over the spectra using the UKIDSS filters bandpasses. Figure~\ref{fig:seds} illustrates the SEDs of Calar\,21 and~22 together with the data of the L7 dwarf J1256$-$1257b \citep{gauza15}, which were normalized to the $J$-band fluxes of the Calar sources. The spectral shapes of Calar\,21 and~22 closely match the SED of J1256$-$1257b, except for the $K$-band, where the latter object is overluminous, which is consistent with its redder $J-K$ color. 

The solar metallicity BT-Settl model atmospheres (\citealt{baraffe15}; see also \citealt{manjavacas16}) were compared to the observed SEDs to assess the most likely $T_{\rm eff}$ and gravity (log\,$g$) parameters. These models self-consistely calculate the thermal structure and equilibrium chemistry for cloud-bearing atmospheres with solar abundance \citep{caffau11} at different resolutions in a wide range of wavelengths. The models also account for the formation and gravitational settling of dust grains at low temperatures. We identified the best spectral fits by visual inspection of the comparisons, which included theoretical spectra covering the following intervals: 1400--1100 K and 2.0--5.0 [cm\,s$^{-2}$] with steps of 100 K and 0.5 dex in $T_{\rm eff}$ and log\,$g$, respectively. Hereafter, models are denoted by the doublets $T_{\rm eff}$/log\,$g$.

We found that Calar\,21 and~22 are reasonably reproduced by the 1300/2.0 BT-Settl model, while Calar\,25's photometric SED is well matched by the 1200/2.0 model. Results are shown in Figure~\ref{fig:models}. Regarding Calar\,25, all photometric data points are reproduced to better than 1\,$\sigma$ the quoted uncertainties. Of the explored grid of models, the 1300/2.0 synthetic spectrum displays the reddest colors, i.e., the steepest rising slope between visible wavelengths and the $K$-band. At warmer and cooler temperatures, the $J-K$ colors are bluer. The BT-Settl models point to that $\sim$1300 K is the temperature at which the settlement of the grains commences for low gravity atmospheres. The derived $T_{\rm eff}$s for Calar\,21, 22, and~25 are consistent with those estimated from evolutionary models (Section~\ref{sec:mass}). We assigned an error of $\pm$100 K to the temperatures estimated from the spectral fitting technique.

However, the spectral fitting analysis yielded a surprisingly low surface gravity that is about two orders of magnitude lower than predicted by the 120-Myr isochrone \citep{baraffe15} depicted in Figures~\ref{fig:zk} and~\ref{fig:hr} (see Section~\ref{sec:sequence}). This low gravity result is driven by the very red colors of the Calar sources. Very low gravities have an effect on the objects' size and mass: for a given mass, the radius increases by a factor or 10, or the mass decreases by a factor of 100 for a fixed size. No evolutionary model available to us predicts such a low gravity for any substellar object with a mass in the interval 0.5--72 M$_{\rm Jup}$ at the age of the Pleiades cluster. The enormous disagreement in the gravity derivations indicates that some physics of substellar evolution is unaccounted for, or that the structure of the dusty atmospheres is poorly determined by the theory, or that the condensate chemistry and contribution of dust to the atmospheres are not properly treated in the computations, or all together. Note that the measured radii of old brown dwarfs by the techniques of transit duration and orbital velocity of double-lined eclipsing binaries (see \citealt{filippazzo15} for a compilation of data) agree with the evolutionary models for typical ages of Gyr (e.g., \citealt{littlefair14}).

Yet, the BT-Settl model atmospheres can be used to determine the trend of various atomic and molecular features with gravity and temperature in qualitatively terms. For a given low temperature, the lower the gravity, the steepest the spectral slope from visible through $\sim$2 $\mu$m. In addition, methane absorption at 2.2 and $\sim$3.4 $\mu$m, which is present in high-gravity 1200-K field dwarfs (early-T types, e.g., \citealt{vrba04}), is clearly sensitive to gravity: methane absorption considerably reduces at low gravity. Oxides have an opposite behavior: they become stronger with lower gravity. A detailed inspection of Figure~\ref{fig:models} reveals that some optical and near-infrared TiO and VO absorption occurs at low atmospheric pressures and temperatures, whereas these molecules disappear at high gravities for the same low temperatures. Since the M/L/T types are defined based on the presence and disappearance of certain molecules from the observed spectra (e.g., TiO, VO, CrH, FeH, CO, CH$_{\rm 4}$), this quite likely affects the establishment of a spectral type definiton for very low gravity dwarfs (see \citealt{osorio17} and \citealt{lodieu17}).


\begin{table}
\centering
\caption{Mass, luminosity, $T_{\rm eff}$, and bolometric correction for three Pleiades low mass members.}
\label{tab:bc}
\resizebox{\columnwidth}{!}{
\begin{tabular}{lcccccc}
\hline
Object    & Mass           & log $L$/$L_\odot$ & $T_{\rm eff}$    & BC($J$)      & BC($K$)       & BC($W1$)\\
          & (M$_{\rm Jup}$) & (dex)            & (K)             & (mag)         & (mag)         & (mag)    \\
\hline
Calar\,21 & 15$^{+5}_{-3}$ & $-$4.29$\pm$0.10 & 1350$^{+100}_{-80}$ & 0.86$\pm$0.15 & 3.37$\pm$0.10 & 4.62$\pm$0.15 \\
Calar\,22 & 15$^{+5}_{-3}$ & $-$4.33$\pm$0.10 & 1350$^{+100}_{-80}$ & 0.89$\pm$0.15 & 3.36$\pm$0.20 & --            \\
Calar\,25 & 11.5$\pm$0.5  & $-$4.66$\pm$0.10 & 1150$^{+50}_{-100}$ & 1.18$\pm$0.20 & 3.55$\pm$0.25 & 4.57$\pm$0.20 \\
\hline
\end{tabular}
}
\end{table}

\subsection{Luminosities and bolometric corrections}
We estimated the bolometric luminosity for Calar\,21, 22, and~25 using the observed spectra and photometry together with the BT-Settl models that provide the best fits to the data as a bolometric correction for short and long wavelengths not covered by the observations. We integrated the extended SEDs over the wavelength interval 3000--10$^6$ \AA~using the simple trapezoidal rule and applied $m_{\rm bol} = -2.5\,{\rm log}\,f_{\rm bol} - 18.9974$, where $f_{\rm bol}$ is in units of W\,m$^{-2}$, to obtain the apparent bolometric magnitudes. They were then converted into absolute bolometric magnitudes using the distance to the cluster of 133.5 pc \citep[and references therein]{galli17}. The solar bolometric magnitude of $M_{\rm bol}$\,=\,$+$4.74 mag was employed to obtain the luminosities of the Calar sources. Results are provided in Table~\ref{tab:bc}. Uncertainties in log\,$L/L_\odot$ were derived from the photometric error bars and the distance error. At the age of the Pleiades, these luminosities correspond to the masses also listed in Table~\ref{tab:bc} and discussed in Section~\ref{sec:mass}. 

\begin{figure}
	\includegraphics[width=\columnwidth]{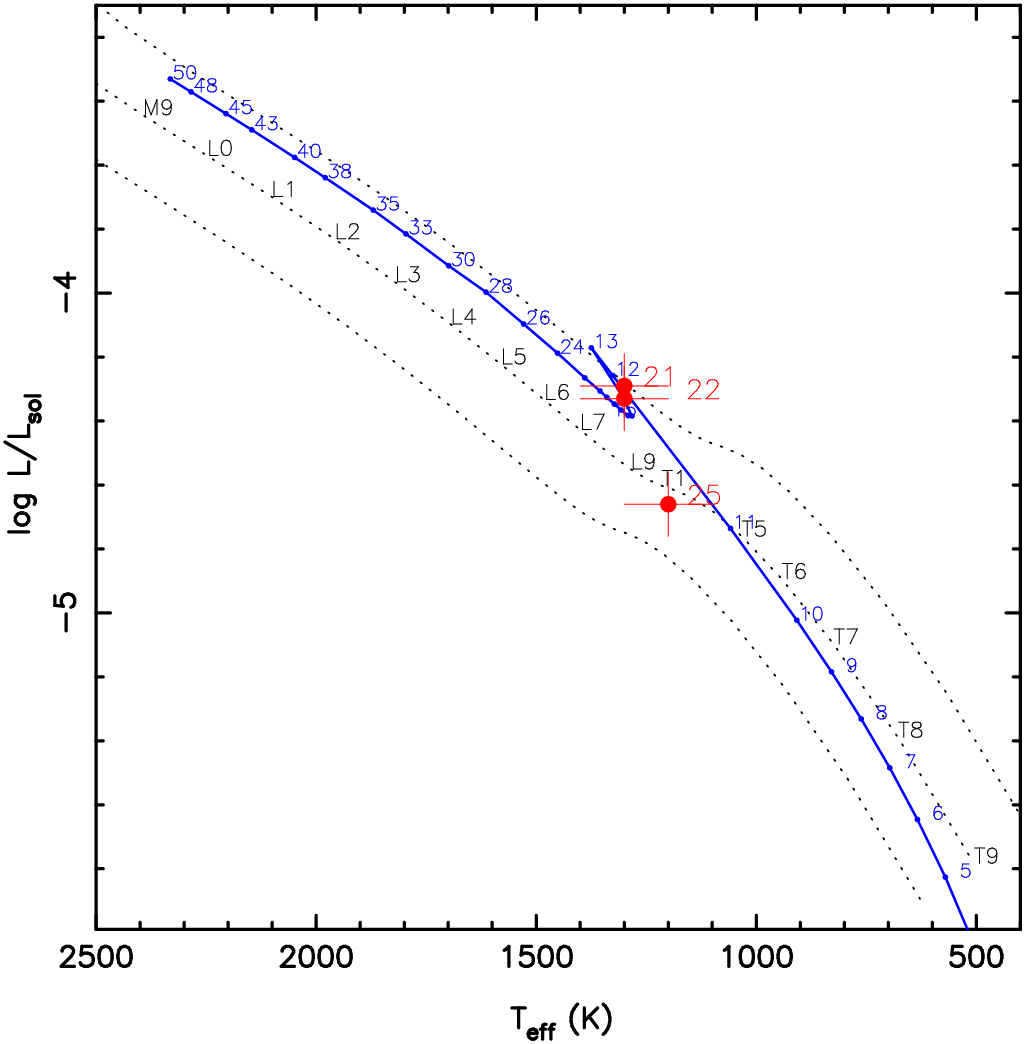}
    \caption{Hertzsprung--Russell diagram of Calar\,21, 22, and~25 (red dots labeled wih the Calar numbers). The field sequence of mid-M through T9 dwarfs defined by \citet{filippazzo15} is indicated with a dotted black line labeled with spectral types. Dotted lines above and below stand for the dispersion observed in the field. The solar metallicity, 120-Myr evolutionary model of \citet{saumon08} is depicted with a solid blue line. Labels in blue correspond to the mass given by the model in Jovian units (1 M$_\odot \approx 1000$ M$_{\rm Jup}$). The kink at about 1300 K correspond to the deuterium burning phase at the age of the Pleiades.
}
\label{fig:lumteff}
\end{figure}

Figure~\ref{fig:lumteff} depicts the Hertzsprung--Russell diagram for Calar\,21, 22, and~25 in comparison with the field sequence of ultracool dwarfs defined by \citet{filippazzo15}. Despite being dimmer in $Z$ and $J$ bands, these Calar objects maintain about the same bolometric luminosity as field dwarfs of similar temperature. Particularly, Calar\,25 has a bolometric luminosity that is undistinguishable from the field. This implies that objects with masses around the deuterium burning-mass limit cool down from the Pleiades age up to Gyr following the track delineated by field late-L and T dwarfs.

The $J$- and $K$-band bolometric corrections (BCs) of Calar\,21, 22, and~25 (Table~\ref{tab:bc}) were obtained from the equation BC($X$) = $m_{\rm bol} - X$, where $X$ stands for the apparent magnitude in each observing filter. Calar\,21 and~22 have BCs that agree with those measured by \citet{filippazzo15} for young dwarfs: BC($K$) is consistent with high gravity dwarfs of related types while BC($J$) appears smaller by about 1 mag with respect to the field (we employed the following references for field BCs: \citealt{dahn02,golimowski04,vrba04,looper08}). This highly contrasts with the results of \citet{todorov10} and \citet{lodieu17}, who found that BC($K$) is larger for the very young Taurus and Upper Scorpius L-type dwarfs. This may be attributable to the different ages ($\sim$1 and $\sim$10 Myr versus 120 Myr) of the samples and to the fact that none of the Taurus and Upper Scorpius sources in those works appear to be as red as our targets. The BCs of Calar\,25, whose spectral type is expected to be equal or cooler than Calar~21 and~22, however, show different properties: BC($K$) is larger than the field (in better agreement with \citealt{todorov10} and \citealt{lodieu17} findings) while BC(J) lies between the values corresponding to young late-L dwarfs of \citet{filippazzo15} and those of the field. These results confirm that BCs of L dwarfs are a function of other parameters than temperature alone: surface gravity, dust concentration, etc.


\subsection{Masses \label{sec:mass}}
Figures~\ref{fig:zk} and~\ref{fig:hr} also depict the BT-Settl 120-Myr isochrone computed for solar metallicity by \citet{allard14} and \citet{baraffe15}. The evolutionary model makes predictions on the behavior of the optical and near-infrared colors as a function of mass (or $T_{\rm eff}$ or luminosity), some of which can be tested against the observed sequence of Pleiades low mass members. The obvious visual result from Figures~\ref{fig:zk} and~\ref{fig:hr} is that the isochrone lacks quantitative accuracy while reproducing the increasing reddening of the cluster sequence from the M through the L types in a qualitative manner. For example, the $Z-K$ and $J-K$ colors appear to saturate at values of $\sim$4.2 and 1.8 mag according to the theory, but Calar~21, 22, and~25 have redder indices by about 1 mag in $Z-K$ (top panel of Figure~\ref{fig:zk}) and about 0.5 mag in $J-K$ (Figure~\ref{fig:hr}); and the $Z-J$ color bends to blue and then to red values, but the theoretical $J$-band magnitude at wich $Z-J$ turns back to red is off by about 0.5 mag with respect to the observed sequence (bottom panel of Figure~\ref{fig:zk}). In the magnitude ranges shown in the Figures, the $Z-K$ index remains ``saturated'' down to very faint magnitudes, whereas the $J-K$ color turns to blue indices caused by the onset of methane absorption and/or the settlement of dust particles below the photospheres. This turnover is predicted to occur at $K \approx 18$ mag at the age and distance of the Pleiades by the BT-Settl model, i.e., a place in between the positions of Calar\,22 and Calar\,25. 

To determine masses, we fixed the age and metallicity of our targets to 120 Myr and solar abundance, respectively. Then, masses were directly obtained from the comparison of the derived bolometric luminosities against the luminosity predictions made by the BT-Settl evolutionary model. This method avoids the uncertainties associated with the derivation of $T_{\rm eff}$ and with the obtention of magnitudes from computed model atmospheres that do not reliably reproduce the observed spectra. Calar\,21 and~22 have a mass of 12--20 M$_{\rm Jup}$ and define the brown dwarf---planetary-mass transition in the Pleiades cluster. The isochrone also delivers $T_{\rm eff}$  for a given luminosity, which is about 1350 K in the case of Calar\,21 and~22. The fainter and redder source Calar~25 has a cooler $T_{\rm eff}$ of $\sim$1150 K and a smaller planetary mass near 11.5 M$_{\rm Jup}$. Table~\ref{tab:bc} summarizes the results. The errors of the mass and temperature derivations account for the quoted uncertainties in luminosity.  For comparison purposes, the masses and $T_{\rm eff}$'s provided by the BT-Settl 120-Myr isochrone for different magnitudes and filters are also labeled in Figures~\ref{fig:zk} and~\ref{fig:hr}. Our luminosity-based mass and temperature derivations agree at the level of 1\,$\sigma$ with the predictions made for the $K$-band, whereas they deviate significantly from the values inferred from the $J$-band ($J$-band-based masses and temperatures are 30--45\%~smaller). This is likely explained by the errors introduced by the theory of model atmospheres.

The masses indicated in Table~~\ref{tab:bc} are very similar to those obtained using the evolutionary cooling sequences of \citet{saumon08}. In order to illustrate the comparison of our data with other evolutionary models, we plotted the \citet{saumon08} 120 Myr, solar metallicity isochrone in Figure~\ref{fig:lumteff}. We selected the ``hybrid'' cloudy/cloudless model, i.e., the computations of an evolution sequence that models the L/T transition by increasing the sedimentation parameter as a function of decreasing $T_{\rm eff}$. The positions of Calar\,21, 22, and 25 are reasonably well described by this ``hybrid'' model, from which we inferred masses of 12--22 M$_{\rm Jup}$ for Calar\,21 and~22, and 11--12 M$_{\rm Jup}$ for Calar\,25. If Calar\,24 were finally confirmed as a Pleiades member, its mass would be estimated at $\approx$11 M$_{\rm Jup}$. In Figure~\ref{fig:lumteff}, the phase of deuterium burning is revealed by the kink in the isochrone at $\sim$1300 K or 12--13 M$_{\rm Jup}$ at the age of the cluster. 

One intriguing feature in Figure~\ref{fig:lumteff} is the pivotal appearance of the theoretical 120-Myr isochrone with respect to the observed sequence defined by field ultracool dwarfs \citep{filippazzo15}: whereas the model is overluminous at warm temperatures, as expected for young ages, it runs below the field sequence for $T_{\rm eff} \le 1100$ K suggesting that Pleiades very low mass members might appear underluminous with respect to the field. We caution that this is not the correct interpretation. Evolutionary models do not predict this underluminosity nature: isochrones of older ages are systematically less luminous than younger isochrones for all masses. The \citet{saumon08} isochrones computed for ages of Gyr do not accurately reproduce the field sequence. This may be caused by the way the average field sequence is obtained (mixing objects of very different ages and metallicities), or by an error in the determination of bolometric luminosities and temperatures in the very low mass regime, or by physics unaccounted for in the evolutionary models. 

Calar\,21, 22, and~25  are excellent targets for carrying out the test of deuterium \citep{bejar99} that has been proposed to discriminate brown dwarfs and planetary-mass objects. All three objects are coeval and have the same metallicity, and their masses are estimated slightly above and below the deuterium burning mass limit at 12--13 M$_{\rm Jup}$ \citep{burrows93,chabrier00b,saumon08}. Calar\,25 is currently the coolest, least luminous and least massive object whose membership in the Pleiades is astrometrically and photometrically confirmed using data from optical to infrared wavelengths. Its finding demonstrates that despite their very low mass and current dynamically relaxed state of the cluster \citep{jameson02,osorio14a}, these objects have not completely escaped the Pleiades.


\section{Conclusions}
\label{sec:conclusions}

This paper presents the follow-up optical imaging and spectroscopy of the six faintest Pleiades candidate members announced in \citet{osorio14a}: Calar 21--26. All data were acquired with the OSIRIS instrument of the GTC. The targets have $J$-band magnitudes in the interval 20.2--21.2 mag; both $Z$-band photometry and low resolution spectroscopy ($R \approx 270$, 6400--10150 \AA) were obtained for Calar~21 and~22 ($J \sim 20.2$ mag), while the fainter sources were observed in the $Z$-band only (except for Calar~24, which was also imaged in the $i$-band). From the imaging analysis, we determined that Calar 23 is resolved and that Calar 26 does not have a proper motion compatible with the Pleiades motion. Therefore, these two objects are regarded as likely non-members of the cluster. The remaining four candidates have new proper motion determinations that agree with the values published in the original paper at the 1 $\sigma$ level and are fully compatible with the Pleiades motion. Calar~21, 22, and~25 show very red $Z-J$, $Z-K$, and $J-K$ colors that overlap with the colors of the reddest known L dwarfs in young stellar moving groups and the field. Despite having similar brightness to Calar~25, Calar~24 is not as red, which may suggest that it is a contaminant of the astrometric survey. However, the colors of Calar~24 are compatible with field late-L or early-T dwarfs, and the position of this object in color-magnitude diagrams coincides with the expected location of the T-type sequence of the Pleiades (according to evolutionary models). On the one hand, we conclude that current data cannot rule out that Calar~24 is an unreddened cluster member. Further near-infrared spectroscopy is required to confirm the low gravity nature of this particular object. On the other hand, we also conclude that Calar~25 remains as an astrometric and photometric likely member of the cluster; it is the reddest cluster object at optical-to-infrared wavelengths. Based on their astrometric, photometric, and spectroscopic properties, Calar~21 and~22 are bona-fide Pleiades members. Calar~21, 22, and~25  currently define the least luminous and least massive tail of the Pleiades sequence.

The optical spectra of Calar~21 and~22 were compared to primary L dwarf standards for spectral classification. We determined L6--L7, which is consistent with the classification from near-infrared spectra \citep{osorio14b}. This agreement between optical and near-infrared spectral types contrasts with the findings in other significantly younger ($\le$10 Myr) clusters like Upper Scorpius \citep{lodieu17} and $\sigma$\,Orionis \citep{osorio17}. These types are expected for Pleiades members with $J \sim 20.2$ mag. At the resolution and quality of the data, there are no strong signatures of TiO and VO absorption and the following features were identified: a markedly rising pseudo-continuum towards longer wavelengths, which is consistent with very red colors, potassium resonance doublet and a tentative detection of neutral cesium at 8521 \AA, and absorption due to ferrum and chromium hydrides. Despite the intermediate-gravity nature of the atmospheres expected for Pleiades low-mass dwarfs (120 Myr), these features show intensities that resemble those of high-gravity dwarfs of similar classification. The comparison with BT-Settl model atmospheres \citep{baraffe15} allowed us to derive effective temperatures of $\sim$1300 K for Calar~21 and~22, and $\sim$1200 K for Calar~25 (error bar of $\pm$100 K), which are close to those obtained from 120-Myr isochrones (1350 and 1150 K, respectively). The bolometric luminosities of Calar~21, 22, and~25 were determined from the integration of their corresponding observed spectral energy distributions, which allowed us to measure bolometric corrections at near-infrared wavelengths for low luminous Pleiades members. These bolometric corrections differ from those of high gravity dwarfs of related classification in the line indicated by \citet{filippazzo15} for low gravity atmospheres. Interestingly, they also differ from the bolometric corrections of much younger L-type sources as measured by \citet{lodieu17}. From evolutionary models \citep{chabrier00a,saumon08}, the mass of Calar~21 and~22 is estimated at 12--20 M$_{\rm Jup}$, while the mass of the least massive Pleiades member, Calar~25, is estimated at  11--12 M$_{\rm Jup}$, below the deuterium burning mass limit. Calar~25 thus turns out to be an excellent candidate for the search for deuterium using upcoming facilities, like JWST. Calar~21, 22, and 25 indicate that the extreme reddish nature of L dwarfs is not always associated with ages younger than $\approx$100 Myr.


\section*{Acknowledgements}

We thank the anonymous referee for useful comments and suggestions. Based on observations made with the Gran Telescopio Canarias (GTC), installed at the Spanish Observatorio del Roque de los Muchachos of the Instituto de Astrof\'isica de Canarias, in the island of La Palma. M.R.Z.O., V.J.S.B., and N. L. acknowledges the financial support from the Spanish Ministry of Economy and Competitivity through the projects AYA2016-79425-C3-2-P and AYA2015-69350-C3-2-P.




\bibliographystyle{mnras}
\bibliography{1} 




\appendix


\section{Follow-up data of NPM03}
\begin{figure}
	\includegraphics[width=\columnwidth]{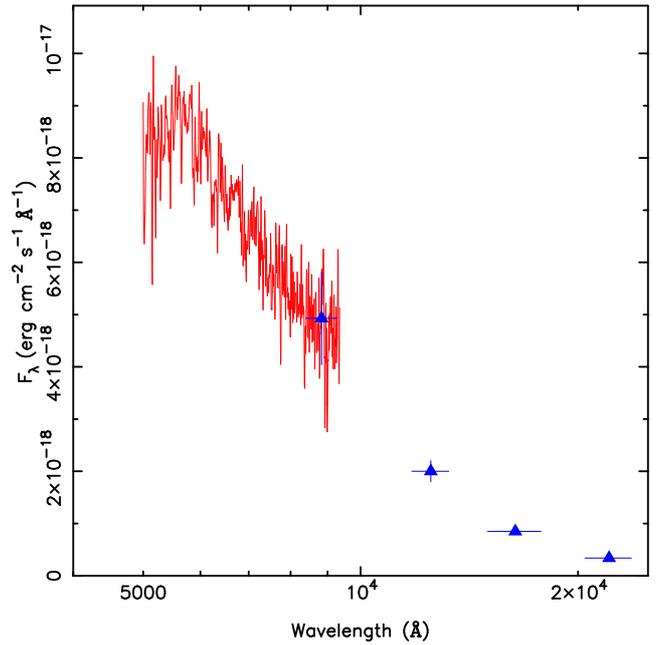}
    \caption{Observed spectrophotometric energy distribution of NPM03. The optical spectrum (red line) and the $Z$-band photometry (blue triangle overlapping with the optical spectrum) are presented here for the first time. The optical spectrum was flux calibrated using the $Z$-band photometry. The horizontal error bars associated with the photometry (blue triangles) indicate the passband of the filters. Near-infrared $JHK$ photometry comes from \citet{osorio14a}. The $x$-axis (wavelength) is shown in logarithmic scale.
}
\label{fig:npm03_sed}
\end{figure}

The new proper motion determination of NPM\,03 is shown in Figure~\ref{fig:pm}. It agrees with that of \citet{osorio14a} within 1 $\sigma$ the quoted astrometric uncertainty. The $Z$-band magnitude (Table~\ref{tab:pm}) indicates that this object is significantly bluer than Pleiades candidate members Calar\,21, 22, and~24 despite having similar $J$-band brightness. In the discovery paper, and prior to the obtention of the data presented here, we especulated that it could be a metal-deficient ultracool dwarf or a white dwarf. The OSIRIS spectrum of NPM\,03 shown in Figure~\ref{fig:npm03_sed} clearly indicates that it is not an M or L or T dwarf. The optical spectrum peaks at around 5600 \AA. The complete energy distribution (constructed as indicated in Section~\ref{sec:models}) suggests that NPM\,03 is a white dwarf. 


\bsp	
\label{lastpage}
\end{document}